# Kinetic roughening and nontrivial scaling in the Kardar-Parisi-Zhang growth with long-range temporal correlations

Tianshu Song[1,2], Hui Xia[1,*]

1) School of Materials Science and Physics, China University of Mining and Technology, Xuzhou 221116, China

2) School of Information and Control Engineering, China University of Mining and Technology, Xuzhou 221116, China

[*] Corresponding author: hxia@cumt.edu.cn

**Abstract:** Long-range spatiotemporal correlations may play important roles in non-equilibrium surface growth process. In order to explore the effects of long-range temporal correlation on dynamic scaling of growing surfaces, we carry out extensive numerical simulations of the (1+1)- and (2+1)-dimensional Kardar-Parisi-Zhang (KPZ) growth system in the presence of temporally correlated noise, and compare our results with previous theoretical predictions and numerical simulations. We find that surface morphologies are obviously affected with long-range temporal correlations, and as the temporal correlation exponent increases, the KPZ surfaces develop gradually faceted patterns in the saturated growth regimes. Our results show that the temporal correlated KPZ system displays evidently nontrivial dynamic properties when $0 < \theta < 0.5$, the characteristic roughness exponents satisfy $\alpha < \alpha_s$, and $\alpha_{loc}$ exhibiting non-universal scaling within local window sizes, which differs with the existing dynamic scaling hypotheses, both in the (1+1)- and (2+1)-dimensions.

## Contents

# 1. Introduction

The Kardar-Parisi-Zhang (KPZ) equation is one of the most widely studied models in the area of surface growth [1], not only represents an extremely important universality class for kinetic roughening, but also can be extended to other research fields [2], such as quantum Bosonic system [3], active matter [4], biological systems [5], and so on. The ($d$+1)-dimensional KPZ equation has the form [1]

$$\frac{\partial h(\boldsymbol{x},t)}{\partial t} = \nu\nabla^2 h + \frac{\lambda}{2}(\nabla h)^2 + \eta(\boldsymbol{x},t), \quad (1)$$

where $h(\boldsymbol{x},t)$ is the growth height at position $\boldsymbol{x}$ and time $t$, $\nu$ is the diffusion coefficient, and $\lambda$ is nonlinear coefficient representing lateral growth. Most stochastic growth models mentioned previously are local, and the noise term $\eta(\boldsymbol{x},t)$ is usually considered as Gaussian distributed, which satisfies the following form,

$$\langle\eta(\boldsymbol{x},t)\eta(\boldsymbol{x}',t')\rangle = 2D\delta^d(\boldsymbol{x}-\boldsymbol{x}')\delta(t-t'), \quad (2)$$

where $d$ is the substrate dimension, the coefficient $D$ represents the noise amplitude.

However, the possibility that the noise has correlations in space and time should be considered in an actual physical system. When the noise has long-range temporal correlations, the second moment (2) is replaced by

$$\langle\eta(\boldsymbol{x},t)\eta(\boldsymbol{x}',t')\rangle = 2D\delta^d(\boldsymbol{x}-\boldsymbol{x}')|t-t'|^{2\theta-1}, \quad (3)$$

where $\theta$ characterizes the decay of temporal correlations. If the exponent $\theta=0$, the noise is uncorrelated, while $\theta=1$ for Brownian motion [6].

Theoretical predictions on the temporal correlated KPZ system are evidently inconsistent with each other based on different analytical approximations, including dynamic renormalization group (DRG) [7], Flory-like scaling approach (SA) [8], and self-consistent expansion (SCE) [9]. Recently, numerical simulations on the KPZ equation with temporal correlations have been performed in (1+1) dimensions, and some results have been obtained correspondingly [10,11]. It should also be noted that there exists a certain differences between the numerical and theoretical predictions, and the scaling results are still unclear. Furthermore, the numerical results in $d>1$, especially the most physically relevant (2+1) dimensions, are still rare when long-range temporal correlations are introduced, and only few theoretical predictions are provided based on the non-perturbative renormalization group (NPRG) approach [12]. Therefore, further exploration of the KPZ equation in the presence of long-range temporal correlations is required. In this article, we revisit the temporal correlated KPZ growth in (1+1) dimensions, and reveal nontrivial dynamic scaling in comparison with the previous results [7-10, 13-15]. Then, we investigate scaling properties in (2+1) dimensional case, and show the comparisons with the theoretical predictions.

The remainder of this paper is organized as follows. First, we briefly present the concepts and dynamic scaling properties of characteristic quantities. Then, we introduce the method to generate desired long-range correlated noise. After that, we show our simulated results in (1+1)- and (2+1)-dimensions, and present the corresponding discussions. Finally, a brief conclusion is given.

## 2. Methods and basic concepts

To study how temporally correlated noise affects scaling properties of the KPZ growth, it needs to generate an effective correlated sequence. In this work we adopt a fast fractional Gaussian noise (FFGN) method originally proposed by Mandelbrot [16], and applied to the temporal correlated Ballistic Deposition (BD) model by Lam et al.[14], which can be briefly described as follows:

$$\begin{cases} B=2, a=6, u_n=aB^{-n}, r_n=e^{-u_n}, \\ W_n^2 = \dfrac{12}{\Gamma(2-2\theta)}(1-r_n^2)(B^{1/2-\theta}-B^{\theta-1/2})(aB^{-n})^{1-2\theta}, \\ X_1(u)=[\zeta_1(u)-1/2]/\sqrt{1-r^2}, \quad for\ t=1, \\ X_t(u)=rX_{t-1}(u)+[\zeta_t(u)-1/2], \quad for\ t=2,3,4,..., \end{cases} \quad (4)$$

where $\zeta(u)$ is uncorrelated noise and uniformly distributed in the interval [0, 1]. The desired correlated noise $\eta_\theta(t)$ can be obtained by $\eta_\theta(t)=\sum_{n=1}^{N}W_n X_t(u_n)$, where $N$ is chosen in the value range of [20, 45] relying on the sequence length we used. Then the correlation function satisfies the following form,

$$C(\tau)=<\eta_\theta(t)\eta_\theta(t+\tau)>\sim \tau^{2\theta-1}. \quad (5)$$

Usually surfaces and interfaces of many growth processes are self-affine, the roughness of surface fluctuations is often quantified by the global surface width $W(L,t)$, defined as the standard deviation of the surface height, which exhibits the following Family-Vicsek (FV) scaling [17],

$$W(L,t)=\sqrt{\sum_{i=1}^{L}[h(\mathbf{x},t)-\bar{h}(t)]^2/L}\sim L^\alpha f(t/L^z). \quad (6)$$

Here, the scaling function $f(u)\sim u^\beta (u<<1)$ and $f(u)\sim const.(u>>1)$, $\alpha,\beta,z$ are roughness, growth and dynamic exponents, respectively.

In addition to the global surface width, the equal-time height-height correlation function $G(l,t)$ is also a commonly used characteristic quantity to investigate scaling properties, which is defined as $G(l,t)=\langle(h(\mathbf{r}+\mathbf{l},t)-h(\mathbf{r},t))^2\rangle$, where $\langle\cdots\rangle$ denotes ensemble average. The function exhibits the scaling form $G(l,t)\sim l^{2\alpha_{loc}}$ ($l<<L$), with $\alpha_{loc}$ denoting the local roughness exponent. The relation $\alpha=\alpha_{loc}$ represents normal FV scaling, and $\alpha\neq\alpha_{loc}$ implies anomalous scaling in surface growth [18].

Furthermore, the structure factor $S(k,t)=\langle\hat{h}(k,t)\hat{h}(-k,t)\rangle$ also plays a major role in exhibiting the behavior of surface growth, where $h(k,t)$ is the Fourier transformation of $h(\mathbf{x},t)$. For ($d+1$)-dimensional stochastic growth models, the structure factor satisfies the scaling form [19, 20],

$$S(k,t) = k^{-(2\alpha+d)} s(kt^{1/z}),  \qquad (7)$$

with

$$s(u) \sim \begin{cases} u^{2(\alpha-\alpha_s)}, & u \gg 1, \\ u^{2\alpha+d}, & u \ll 1, \end{cases}$$

where $s(u)$ is the spectral function, and $\alpha_s$ denotes the spectral roughness exponent. The surface roughening follows normal FV scaling when $\alpha_s = \alpha$, and anomalous roughening appears when $\alpha \neq \alpha_s$. To determine the universal scaling behavior, we adopt the scaled spectral function $S(k,t)k^{(2\alpha+d)}$ versus the scaled wave number $kt^{1/z}$. Therefore, we obtain the scaling exponents $\alpha$ and $z$ based on data collapse. And we also check the dynamic exponent obtained independently based on the scaling relation $z = \alpha/\beta$.

## 3. Results in (1+1) dimensions

In this section, we revisit the temporal correlated KPZ growth in (1+1) dimensions, then investigate the scaling results, and compare with that of the existing numerical studies and analytical predictions. Here, we perform extensively numerical simulations with larger system size and more independent runs, and obtain more scaling exponents in comparison with the previous simulated results [10, 11]. In our simulations, the hybrid computing between graphics processing units (GPUs) and CPUs is performed extensively to improve the computing efficiency.

When performing the simulations, we notice that numerical divergence always exists in the discretized version of the temporal correlated KPZ equation in the effective $\lambda$ regime. It is impossible to follow numerically the evolution of the system beyond the time threshold when this singular growth occurs. In order to control numerical instability, we adopt the scheme through replacing the nonlinear term by an exponentially decreasing function, as suggested by Dasgupta et al. [21],

$$f(x) \equiv (1-e^{-cx})/c, \tag{8}$$

where $c$ is an adjustable parameter. Thus the improved temporal correlated KPZ equation in (1+1) dimensions is discretized as

$$\frac{h(x,t+1)-h(x,t)}{\Delta t} = \frac{\nu}{(\Delta x)^2}[h(x+\Delta x,t)-2h(x,t)+h(x-\Delta x,t)] \\ + \frac{\lambda}{2}[1-e^{-c[\frac{h(x+\Delta x,t)-h(x-\Delta x,t)}{2\Delta x}]^2}]/c + \eta_\theta(x,t). \tag{9}$$

Here, we choose parameters $\nu=1$, $\lambda=4$, $c=0.1$, and the spatial and temporal steps $\Delta x=1$ and $\Delta t=0.05$, respectively. The system is started from a flat interface at $t=0$ with periodic boundary conditions.

To investigate how long-range temporal correlations affects the surface morphology, we show the morphology evolution with different temporal correlation exponents at the early and saturated growth regimes. Figure 1 shows the interface profiles of the KPZ system affected obviously by temporally correlated noise. Figure 1(a)-(d) shows the typical interface profiles with $\theta=0.05, 0.20, 0.40, 0.48$ at early growth time. Similarly, Figure 1(e)-(h) exhibits the interface profiles in the saturated growth regime with the same temporal correlated exponents chosen in Figure 1(a)-(d). We find that, from the early growth to the saturation growth time, the KPZ system with small temporal correlated noises exhibits self-affine interface. However, when temporal correlation exponent increases continually, self-affine structures will be broken, and the interfaces finally develop gradually faceted patterns in the saturated growth regimes. The obvious segmentation happens when the temporal correlation exponent is beyond a certain critical threshold [11]. We also notice that, although there exist similar evolving trends, the interface profiles in our work differs slightly with the previous simulations [11, 13]. Specially, there exists very small spears in simulating KPZ and BD driven by long-range temporally correlated noise [11], and these special small structures do not appear in our simulations in (1+1)-dimensions. With temporal correlations increasing and time reaching the saturated growth regimes, interfaces gradually form similar mounded structures on different scales. Remarkably, when $\theta$ is close to 0.5, the temporal correlated KPZ interfaces develop full faceted patterns in the long growth time limits.

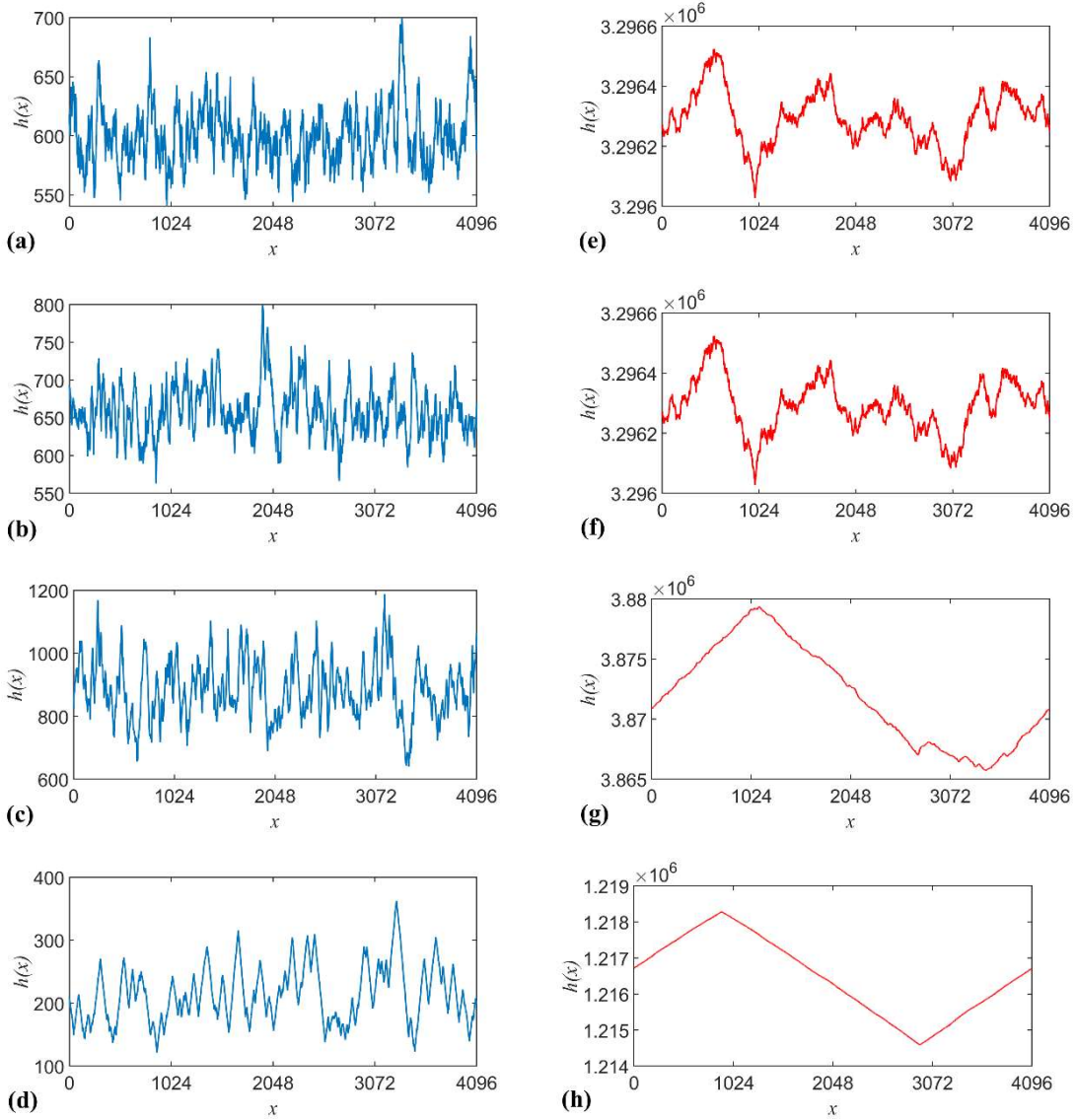

Figure 1. The morphology of KPZ equation in (1+1) dimensions at the early growth time $t = 1.28 \times 10^3$ with $(a)\theta=0.05, (b)\theta=0.20, (c)\theta=0.40, (d)\theta=0.48$, and at the saturated growth time $t = 4.81 \times 10^6$ with $(e)\theta=0.05, (f)\theta=0.20, (g)\theta=0.40, (h)\theta=0.48$.

In order to obtain scaling exponents, we calculate the global surface width $W(L,t)$ based on power-law scaling from Equation (6) in the early and saturated growth regimes separated by a crossover time $t_\times \sim L^z$: $W(L,t) \sim t^\beta$ for $t \ll t_\times$ and $W(L,t) \sim L^\alpha$ for $t \gg t_\times$. Thus, we obtain the effective scaling exponents $\beta$ and $\alpha$ in the early growth and the saturated growth regimes, respectively.

Figure 2 exhibits the log-log plot of $W(L,t)$ versus $t$ with $\theta = 0.20$ and $\theta = 0.40$ as two typical representatives for the small and large temporal correlations, respectively. Here $L = 65536$ is used to increase the precision. In the early time regime, we obtain the growth exponents $\beta=0.41$ for $\theta = 0.20$, and $\beta=0.65$ for $\theta = 0.40$. In the saturated growth regime, we obtain the global roughness exponents $\alpha=0.60$ for $\theta = 0.20$, $\alpha=0.90$ for $\theta = 0.40$, and these results are shown in the insets of Figure 2.

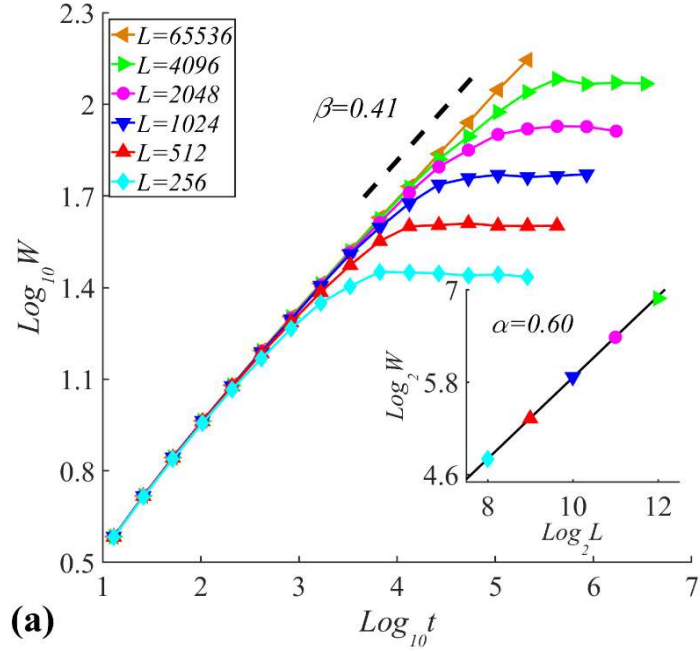

(a)

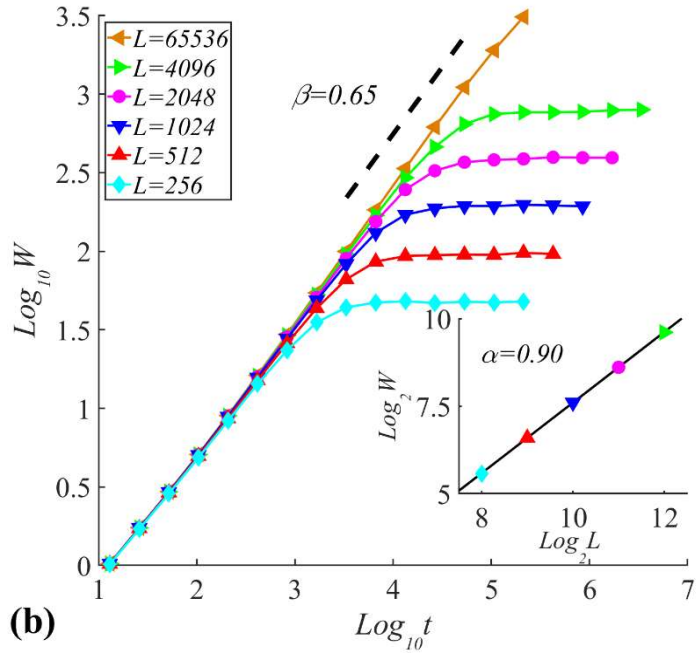

Figure 2. The global surface width $W(L,t)$ versus growth time $t$ with (a) $\theta = 0.20$; (b) $\theta = 0.40$. All results have been averaged over 500 noise realizations, and dashed lines are plotted to guide the eyes. Insets show the saturated surface width $W_{sat}$ and system size $L$, and the solid lines are the fitting values of global roughness exponent $\alpha$.

In order to study further scaling properties, we rescale the surface width by $W/L^{\alpha}$ and rescale the growth time by $t/L^{z}$. According to Equation (6), these curves with the chosen system sizes and growth times collapse to a single curving line based on rescaling, as shown in Figure 3. These results also show that the values of dynamic exponent obtained independently from data collapses are in agreement with those obtained from $z = \alpha/\beta$.

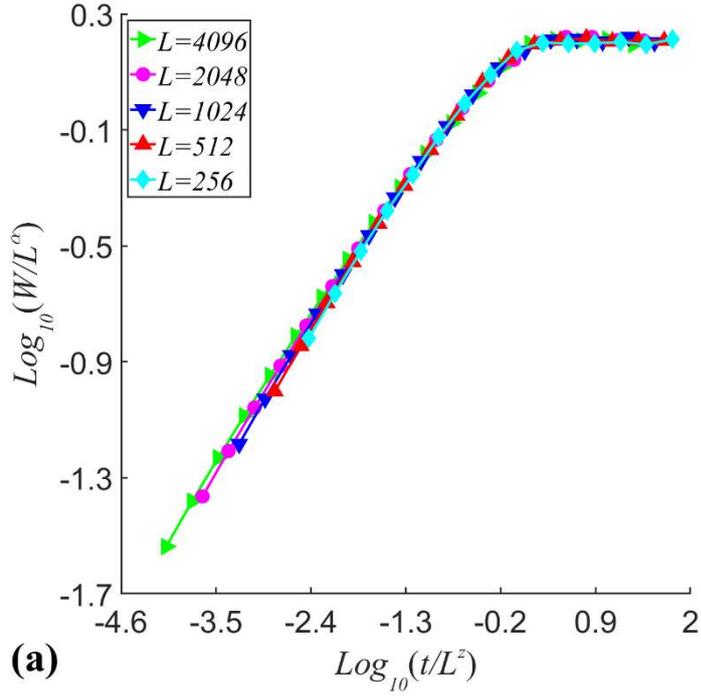

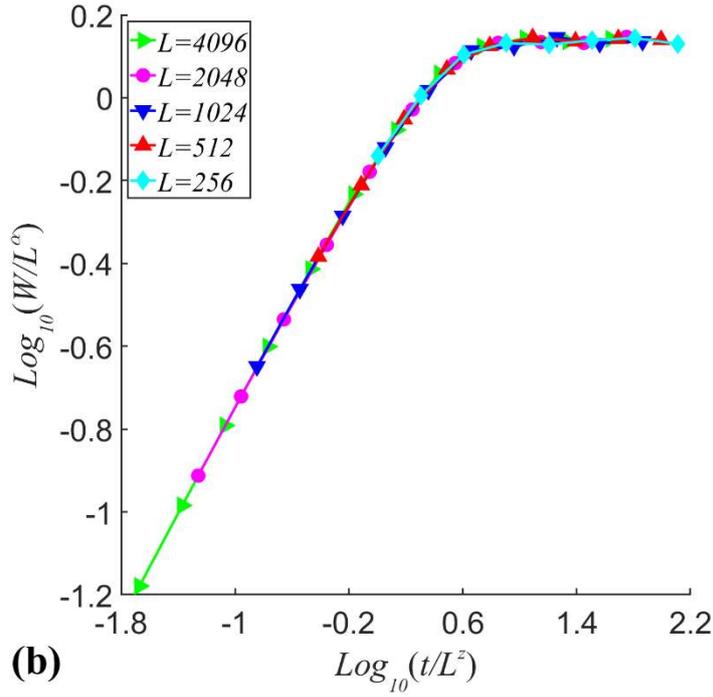

Figure 3. The log-log plot of the scaled global surface width versus the scaled growth time with (a) $\theta = 0.20$; (b) $\theta = 0.40$. Results show data collapses with the chosen scaling exponents: (a) $\alpha = 0.60$ and $z = 1.47$; (b) $\alpha = 0.90$ and $z = 1.37$.

In order to investigate further scaling properties of the KPZ growth, we compute the height-height correlation function $G(l,t)$ with different $\theta$, and obtain $\alpha_{loc}$ using the scaling form $G(l,t) \sim l^{2\alpha_{loc}}$ within the small size, as shown in Figure 4. Here, we choose $L = 32768$, and $2 \leq l \leq 1024$ in order to satisfy the local window size $l \ll L$. Remarkably, we find show that the height-height correlation function does not exhibit universal scaling behavior within local window sizes, and thus the local roughness exponent is not universal, more precisely, the local roughness exponents tend to decrease with window size. Therefore, there are no evidences for the existence of a universal critical exponent describing local window in the KPZ growth driven by the correlated noise.

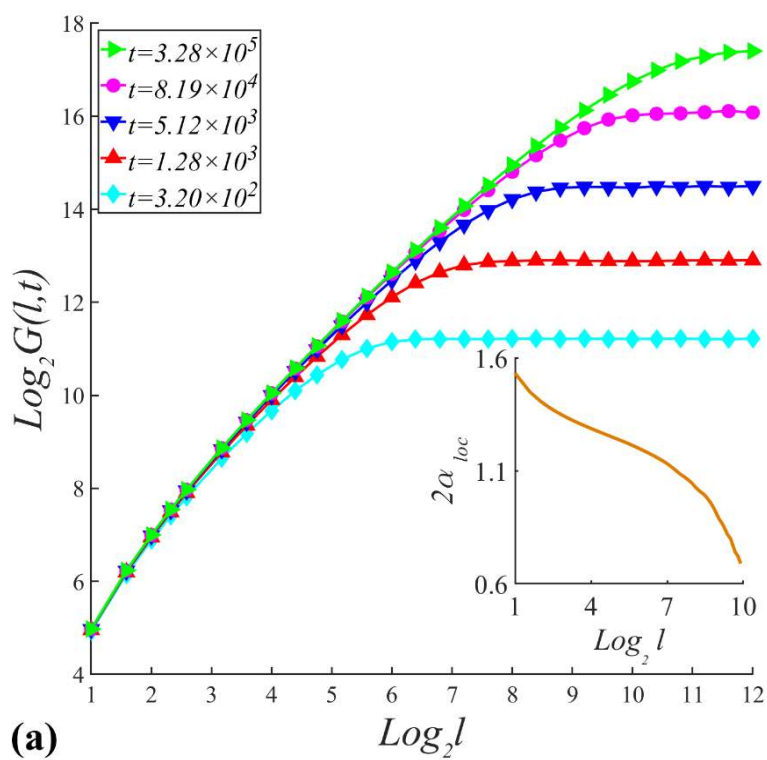

(a)

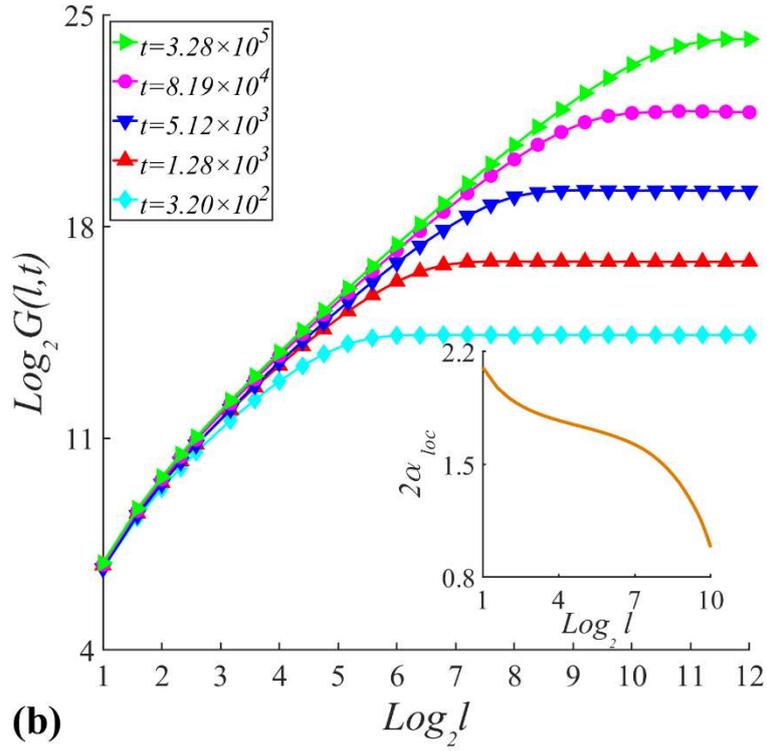

Figure 4. The log-log plot of the height-height correlation function $G(l,t)$ versus window size $l$ with $L=32768$: (a) $\theta = 0.20$; (b) $\theta = 0.40$. Results have been averaged over 500 noise realizations. For clear comparison, each curve shifts accordingly along the vertical coordinate. Insets show $\alpha_{loc}$ versus $l$, which are obtained from the simulations of $t = 3.28 \times 10^5$.

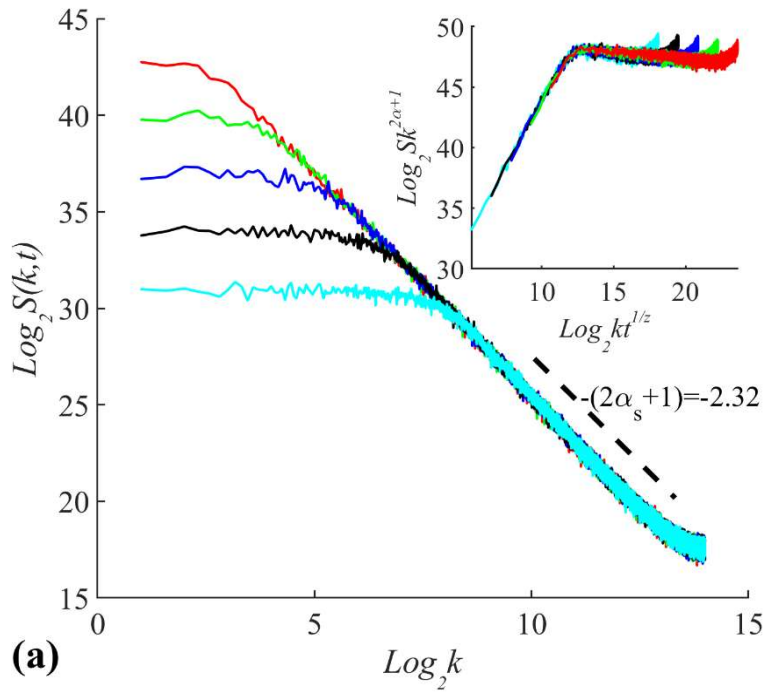

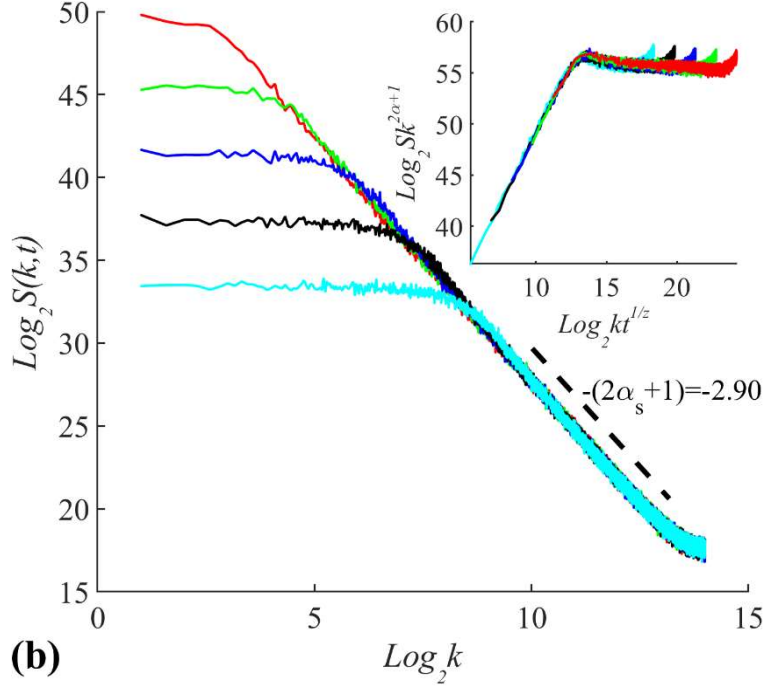

Figure 5. The log-log plot of the structure factor $S(k,t)$ versus wave number $k$ with $L=32768$ and growth times $t=1.28\times10^3, 5.12\times10^3, 2.05\times10^4, 8.19\times10^4, 3.28\times10^5$: (a) $\theta=0.20$; (b) $\theta=0.40$. Results have been averaged over different noise realizations: 200 for $\theta=0.20$ and 150 for $\theta=0.40$. Insets show data collapses with the scaling exponents: (a) $\alpha=0.61, z=1.45$; (b) $\alpha=0.90, z=1.37$.

Figure 5 shows the log-log plot of the structure factor $S(k,t)$ versus wave number $k$ with $\theta=0.20$ and $\theta=0.40$. By performing simulations with different system sizes and growth times, we find that the scaling behavior of the structure factor do not change evidently with $L$ and $t$. Using the scaling relation $S(k,t) \sim k^{-(2\alpha_s+1)}$ in the large wave number regime, we obtain the spectral roughness exponent $\alpha_s$ as an independent critical exponent. In Figure 5(a), we choose $\theta=0.20$ as a typical example in the small $\theta$ regime, and obtain $\alpha_s=0.66$. Based on data collapses as shown in the inset of Figure 5(a), we obtain independently $\alpha=0.61$ and $z=1.45$. In Figure 5(b), we choose $\theta=0.40$ representing a large value of the temporal correlation, and obtain $\alpha_s=0.95$. Meanwhile, we also obtain $\alpha=0.90$ and $z=1.37$ based on data collapse, which is consistent with the values obtained from Equation (6).

We find that the global roughness exponents are less than the corresponding spectral scaling exponents, and the local roughness does not exhibit universal scaling when $0<\theta<0.5$. Thus our results provide numerical evidences that nontrivial scaling can occur even in the small $\theta$ regime, Therefore, these results imply that the temporal correlated KPZ growth system displays nontrivial dynamic scaling, which is a little different with the recent results [11]. It is remarkable that $\alpha<\alpha_s$ does not satisfy the existing dynamic scaling classifications [22].

In Figure 6, we compare the critical exponents $\alpha_s$, $\alpha$, $\beta$ and $z$ with the previous results from analytical predictions and numerical simulations. We find that the spectral roughness exponent is always larger than the global roughness exponent in the presence of long-range temporal correlations, which slightly differs from the previous numerical results [10]. For the global roughness exponents, with comparison of the previous theoretical predictions, we find that these numerical results are consistent with SCE [9] in the small $\theta$ regime, and are in agreement with FRG [15] in the large $\theta$ regime. However, we also find that the growth exponent is not consistent with any existing analytical results for the large $\theta$ regime. Furthermore, our results show that the local roughness exponent is not a universal value, implying that local window sizes do not exhibit scaling in the whole correlation regime. It should be noticed that, the values of the growth and roughness exponents are somewhat larger than the previous results of the BD model in presence of long-range temporal correlations [13, 14]. This discrete growth model has long been regarded as the same universality class with the temporal correlated Kardar-Parisi-Zhang (KPZ) equation. And the dynamic exponent can be obtained using $z=\alpha/\beta$, meanwhile, we also obtain the values of $z$ based on data collapses from the scaled surface width and the scaled structure factor independently. We notice that, although the values of $z$ obtained by these two ways are in line within the whole temporal correlated regime, there also exists a certain difference between them. Considering that the value of $\beta$ is more small and sensitive in comparison with that of $\alpha$, it is possible to enlarge the fluctuations of indirectly measuring $z$. These critical values obtained here are consistent with each other for a fixed value of temporal correlation exponent. We also noticed that, when $\theta$ is near 0.5, the critical values of both $\alpha$ and $\alpha_s$ approach to 1, although not fully understood yet, which implies that there exists an intrinsic relation quantifying the development of the KPZ interface to full faceted patterns in the long growth time limit.

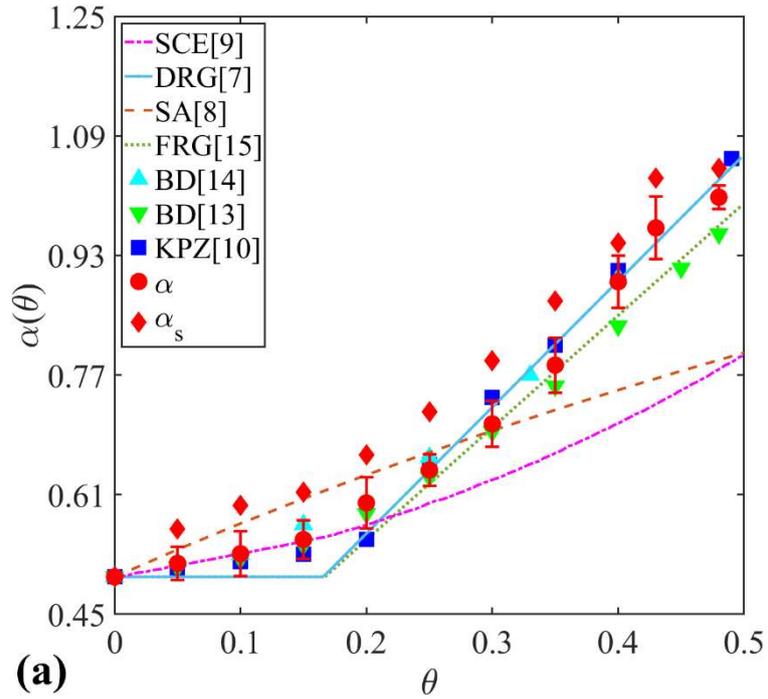

(a)

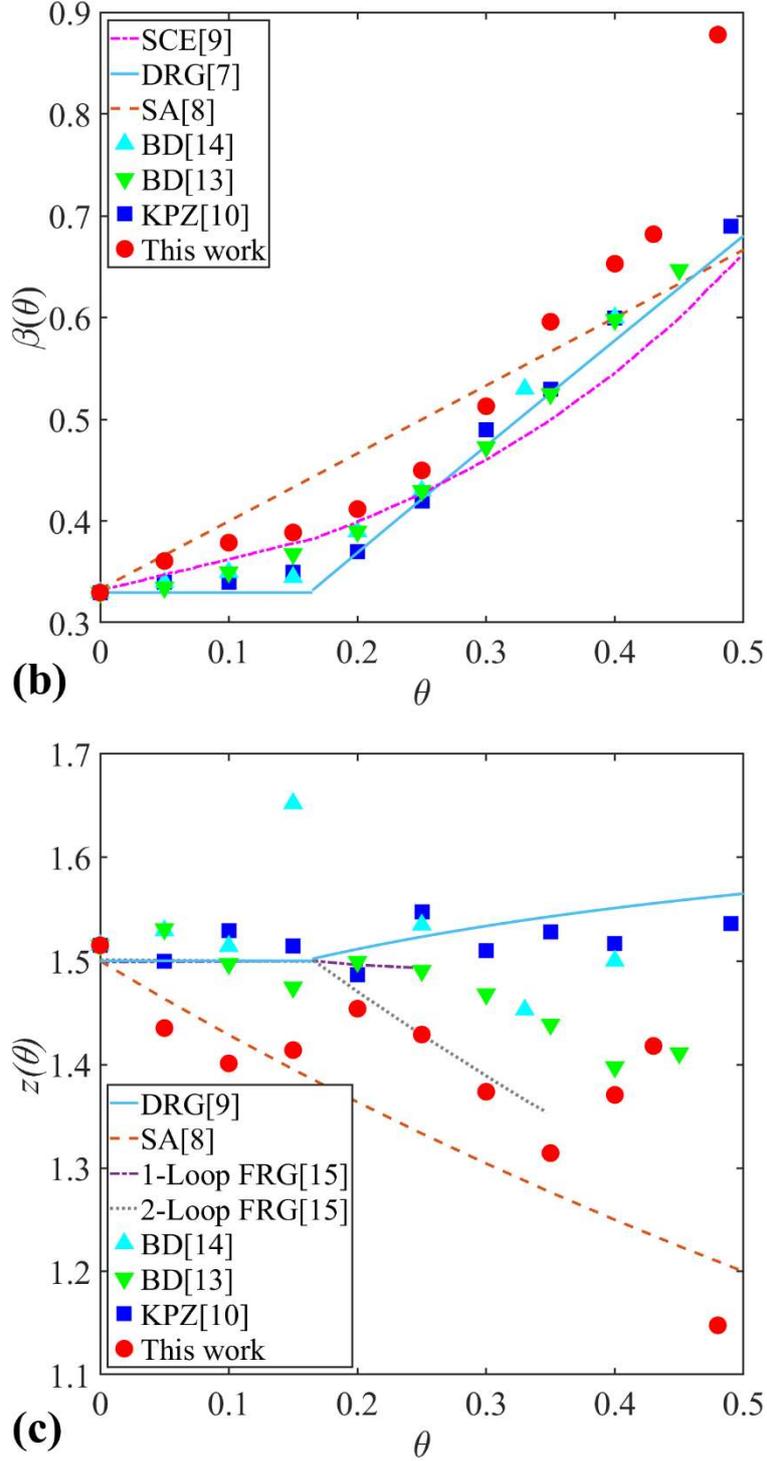

Figure 6. The scaling exponents as a function of $\theta$ for the temporal correlated KPZ system: (a) $\alpha$ versus $\theta$; (b) $\beta$ versus $\theta$; (c) $z$ versus $\theta$. The existing theoretical predictions and numerical results are also provided for a quantitative comparison.

## 4. Results in (2+1) dimensions

For the (2+1)-dimensional case, the temporal correlated KPZ equation with exponentially decreasing function can be discretized as

$$\frac{h(x,y,t+1)-h(x,y,t)}{\Delta t} = \nu[\frac{h(x+\Delta x,y,t)-2h(x,y,t)+h(x-\Delta x,y,t)}{(\Delta x)^2}$$
$$+\frac{h(x,y+\Delta y,t)-2h(x,y,t)+h(x,y-\Delta y,t)}{(\Delta y)^2}]+\eta_\theta(x,y,t) \quad (10)$$
$$+\frac{\lambda}{2}[1-e^{-c\{[h(x+\Delta x,y,t)-h(x-\Delta x,y,t)]^2/4(\Delta x)^2+[h(x,y+\Delta y,t)-h(x,y-\Delta y,t)]^2/4(\Delta y)^2\}}]/c.$$

Here, we set the same parameters with those of the (1+1)-dimensional case, and the spatial discretization $\Delta x = \Delta y = 1$. As a special case $\theta=0$, we first revisit simulating KPZ equation without temporally correlated noise, and our obtained values are in agreement with the existing predictions and simulation results from various discrete and continuum growth models belonging be the KPZ universality class [23-32]. However, numerical investigation on the KPZ system with long-range temporal correlation is still lacking. First, we provide evidences on surface morphology affected by long-range temporal correlation, the temporal correlated KPZ surfaces with different temporal correlation exponents and growth times are illustrated as shown in Figure 7.

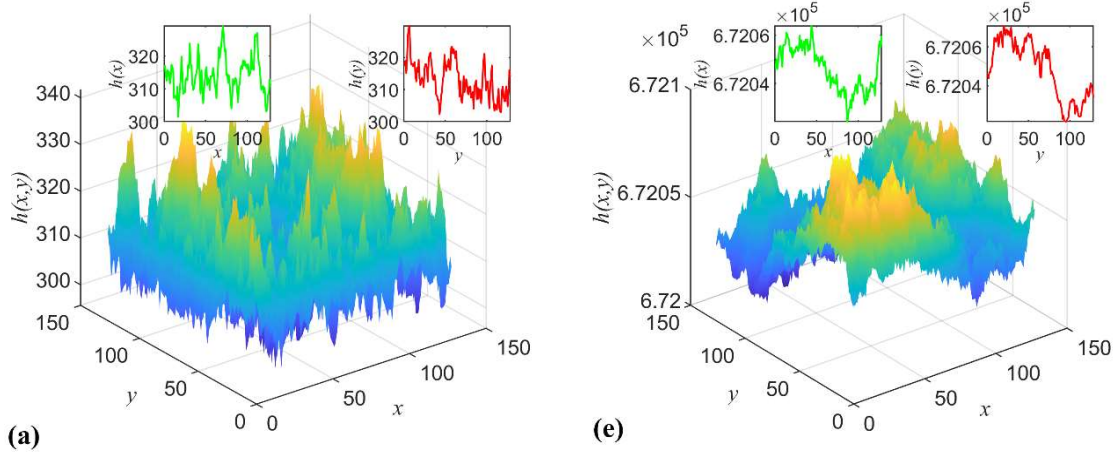

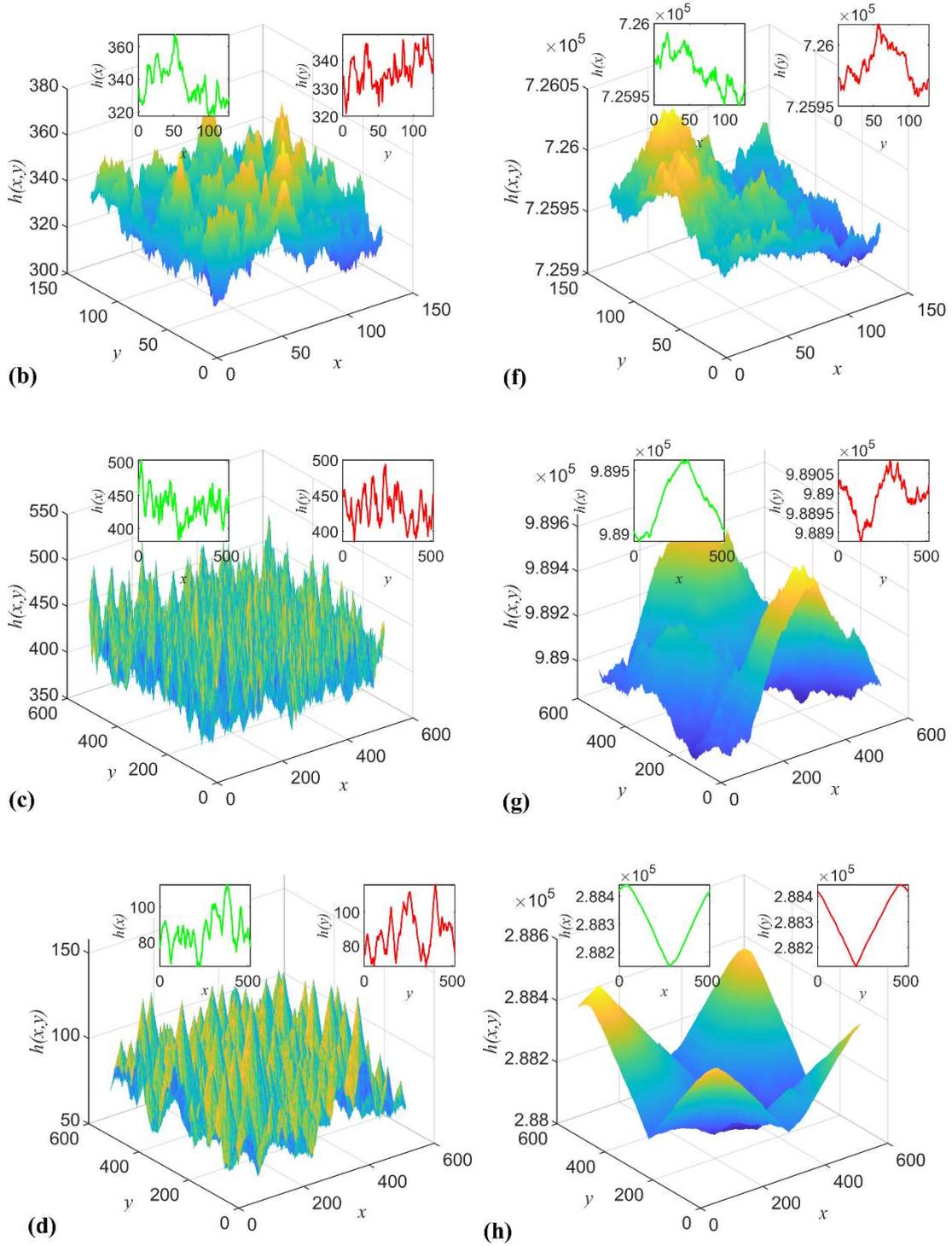

Figure 7. The morphology of the temporal correlated KPZ equation in (2+1) dimensions at the early growth time $t = 6.3 \times 10^2$ with $(a)\theta=0.05, (b)\theta=0.20, (c)\theta=0.40, (d)\theta=0.48$, and at the saturated growth time $t = 1.30 \times 10^6$ with $(e)\theta=0.05, (f)\theta=0.20, (g)\theta=0.40, (h)\theta=0.48$. The subgraphs of each subfigure are interfaces of cross section at $X = 1$ and $Y = 1$.

Figure 7(a)-(d) display surface morphologies at the early growth time with $\theta$=0.05, 0.20, 0.40, 0.48 respectively, and Figure 7(e)-(h) show the corresponding morphologies in the saturated growth regimes. We observe that, when $\theta \leq 0.20$, there is no any obvious mounds observed, however we see the mound-shaped surfaces when $\theta \geq 0.40$. Therefore, there exists a critical threshold $\theta_\tau$ gating two different types of surface morphology. With increasing long-range temporal correlation, the KPZ growth will eventually form the full mounded structure at the long growth time, as shown in Figure 7(h). Furthermore, although faceted patterns have also been observed on the temporal correlated BD model for (2+1) dimensional case [13], through careful comparison, we observe that the evolving morphology exists a certain difference between KPZ and BD when long-rang temporal correlations are introduced, especially for the large temporal correlation and the saturated growth regions.

    For the (2+1)-dimensional case, it is difficult to analyze the temporal correlated KPZ system theoretically. Thus, direct simulating KPZ becomes an effective tool to investigate scaling properties. It should be noted that it takes too much computation time to obtain reasonable results in (2+1)-dimensions. Therefore, the hybrid computing scheme employing both GPUs and CPUs is adopted in order to improve efficiency in our simulations. The global surface width versus the growth time with different system sizes are illustrated in Figure 8. Here, $\theta$=0.20 and $\theta$=0.40 are chosen as two representatives of temporal correlation parameters. When $\theta$=0.20, we obtain $\beta$=0.28 and $\alpha$=0.48, and when $\theta$=0.40, $\beta$=0.59 and $\alpha$=0.82. Therefore, we find that temporal correlated exponents evidently affect evidently the scaling behavior at the whole growth regimes for the (2+1)-dimensional case.

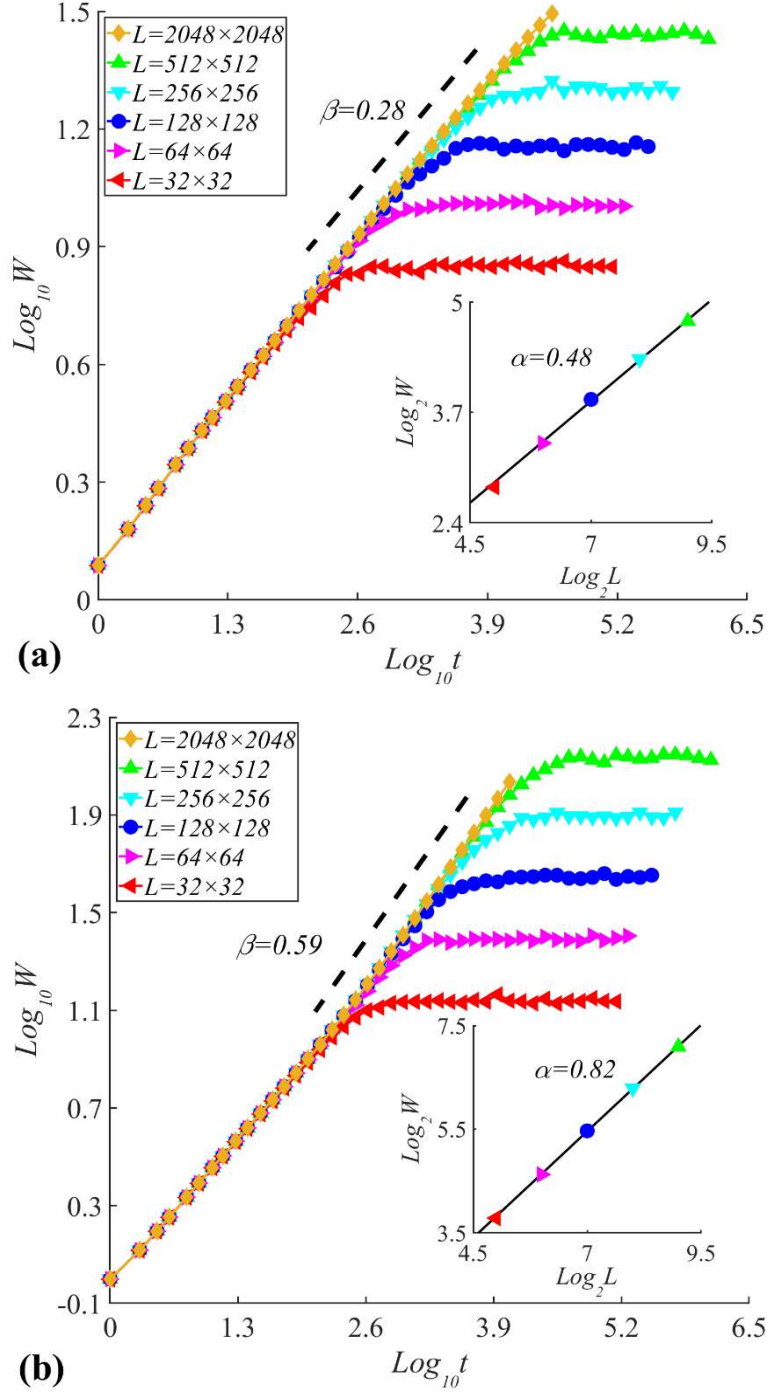

Figure 8. $W(L,t)$ versus $t$ in (2+1) dimensions: (a) $\theta = 0.20$; (b) $\theta = 0.40$. Results have been averaged over 150 noise realizations, and dashed lines are plotted to guide the eyes. Insets are the log-log plot of the saturated surface width and system size, and the solid lines are the fitting results with the values of global roughness exponent.

In the following, we check the scaling exponents based on data collapse. We rescale by $t \to t/L^z, W \to W/L^\alpha$, different curves in Figure 8 collapse into one curve which conforms the scaling hypothesis (6) with properly chosen scaling exponents. The rescaled curves are shown in Figure 9.

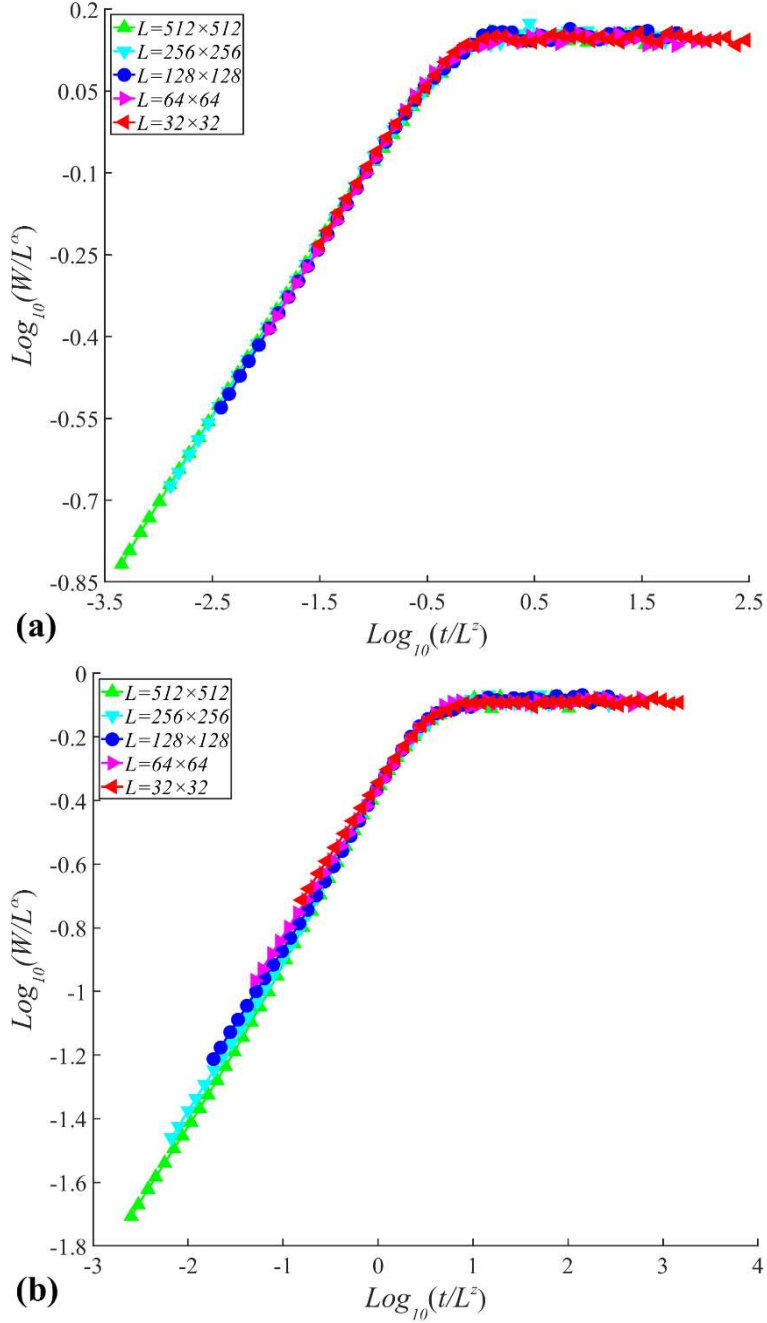

Figure 9. The log-log plot of the scaled surface width versus the scaled growth time: (a) $\theta = 0.20$; (b) $\theta = 0.40$, displaying data collapses with the chosen critical exponents: (a) $\alpha = 0.48$ and $z = 1.73$; (b) $\alpha = 0.82$ and $z = 1.38$.

In order to investigate further scaling properties in (2+1) dimensions, we compute the height-height correlation function $G(l,t)$ with different temporal correlation exponents. Here, we choose $L \times L = 4096 \times 4096$, and $2 \times 2 \leq l \times l \leq 128 \times 128$ in order to satisfy the local window size $l \ll L$. These results are illuminated in Figure 10. Remarkably, similar to the (1+1)-dimensional case, our results show that the height-height correlation function does not exhibit universal scaling behavior with the local window size, and the local roughness exponent is not universal. More precisely, the local roughness exponent sharply decreases with increasing window size, and there is no evidence of universal critical exponent describing local window size. Therefore, the KPZ growth in the presence of temporally correlated noise displays nontrivial scaling behavior in comparison with the local KPZ driven by uncorrelated noise.

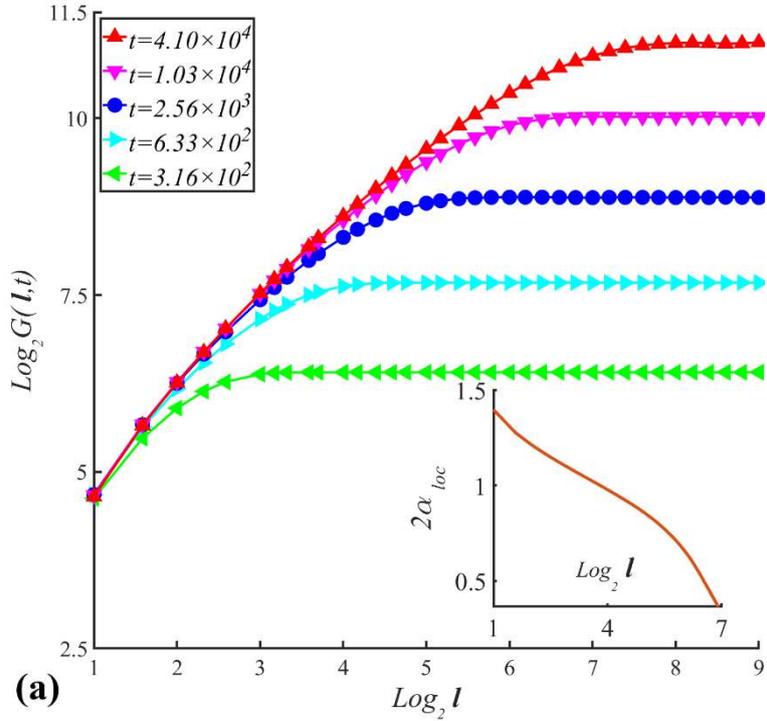

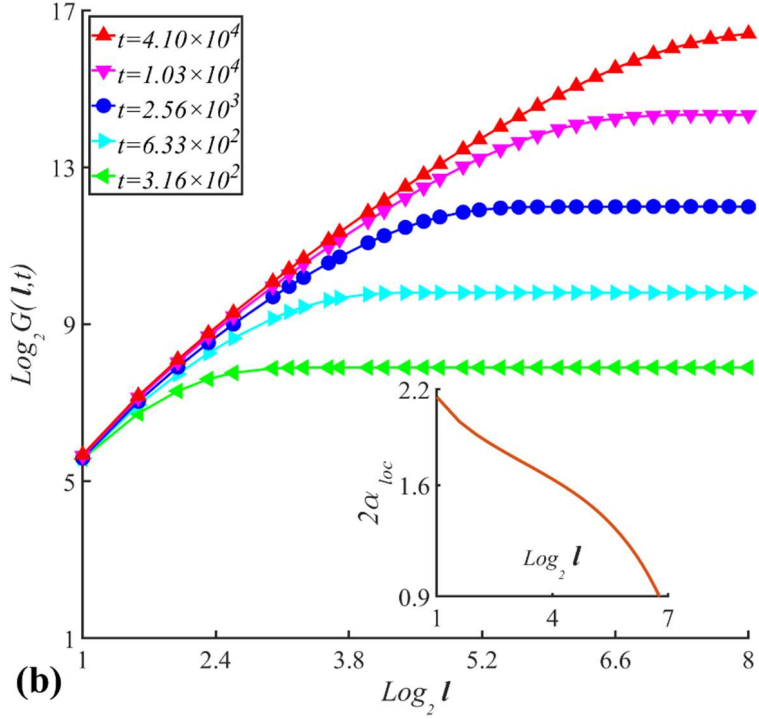

Figure 10. The results of the height-height correlation function with two temporal correlated exponents: (a) $\theta = 0.20$; (b) $\theta = 0.40$. Insets show $\alpha_{loc}$ versus $l$, which are obtained from the simulations of $t = 4.10 \times 10^4$.

In order to further investigate scaling behavior in (2+1)-dimensions, we also calculated the structure factor. The numerical results of spectrum roughness are shown in Figure 11. We obtain $\alpha_s = 0.54$ for $\theta=0.20$, and $\alpha_s = 0.96$ for $\theta=0.40$ from the power-law scaling of the structure factor versus the wave number. Based on data collapses, we also obtain independently the scaling exponents $\alpha$ and $z$ for $\theta=0.20$ and $\theta=0.40$, as shown in insets of Figure 11. We find that, similar to (1+1)-dimensional case, the global roughness exponent is always less than the spectrum roughness when $0 < \theta < 0.5$.

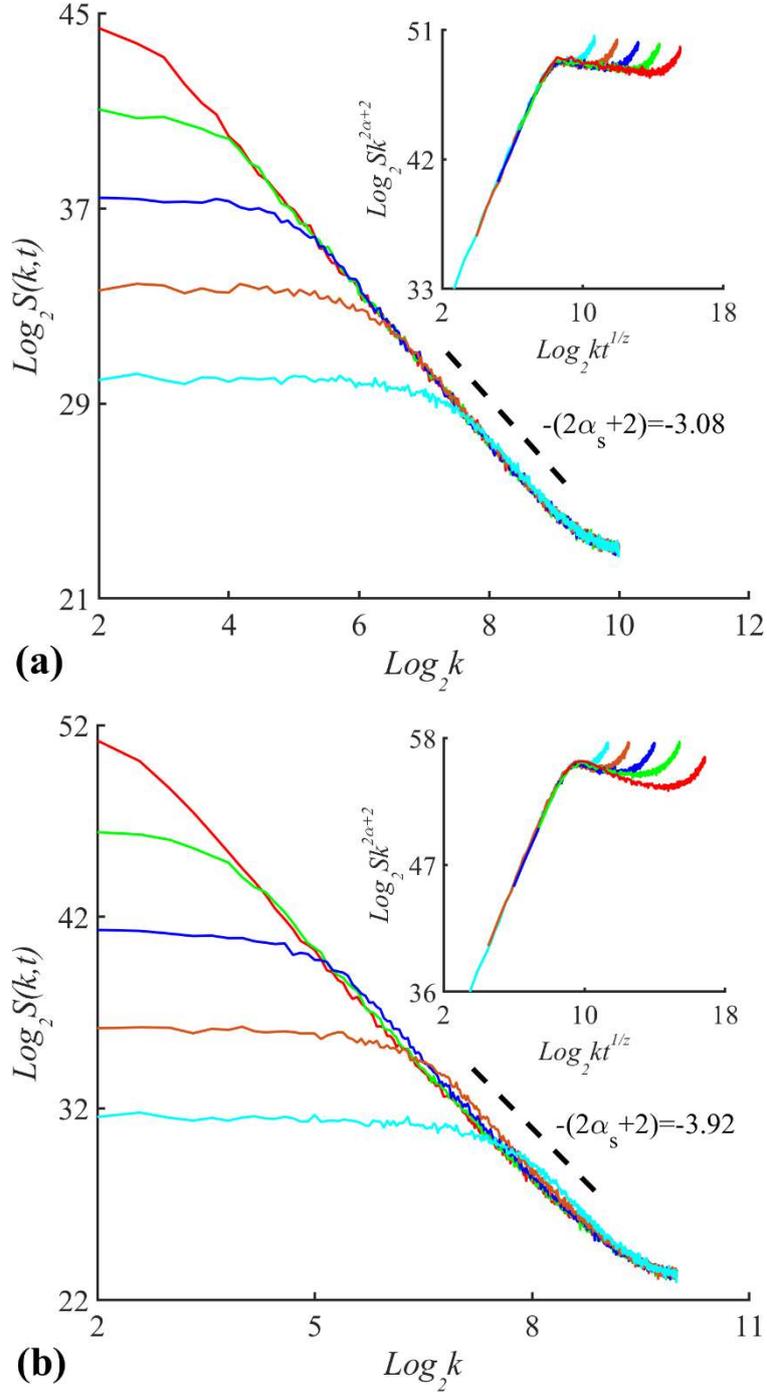

**Figure 11.** The structure factor $S(k,t)$ of the growth surface with different growth times $t = 1.60 \times 10^2, 6.40 \times 10^2, 2.56 \times 10^3, 1.02 \times 10^4, 4.10 \times 10^4$: (a) $\theta$=0.20; (b) $\theta$=0.40. Insets show data collapses with the chosen critical exponents: (a) $\alpha$=0.50 and $z = 1.68$; (b) $\alpha$=0.88 and $z = 1.36$. The system size $L = 2048 \times 2048$ is used, and these data are averaged over 150 independent noise realizations.

To make full comparisons of the scaling exponents obtained, we illustrate the values of $\alpha_s$, $\alpha$, $\beta$ and $z$ in Figure 12. When $\theta > 0$, only few theoretical predictions were reported in the (2+1)-dimensional case. The existing theoretical predictions and numerical results are listed correspondingly. The values of $\alpha$ and $\alpha_s$ are shown in Figure 12(a), and the values of $\beta$ and $z$ are listed in Figure 12(b) and Figure 12(c), respectively. Comparison shows that, our results are consistent with the previous results when $\theta = 0$, however, the results are not in line with the theoretical predictions of one-loop DRG technique [7] and the non-perturbative renormalization group (NPRG) approach [12]. And we also observe that our values of $\beta$, $\alpha_s$ and $\alpha$ are a little larger than those from simulating BD model driven by long-range temporally correlated noise, which is regarded as the same universality class with the temporal correlated KPZ growth in (2+1)-dimensions [13].

We find from Figure12(c) that the values of $z$ are consistent with the predictions of the next-to-leading order (NLO) approximation [12] and one-loop DRG approaches [7] in the small $\theta$ regime, but not so when $\theta > 0.3$. We also notice that, although there exists a certain fluctuations, $z(\theta)$ decreases gradually with $\theta$ increasing in both (1+1)- and (2+1)-dimensional cases, which are also different with the previous results, including KPZ [10] and BD [14] driven by long-range temporally correlated noise. Our results imply that the temporal correlated KPZ system in (2+1) dimensions displays nontrivial dynamic scaling. Remarkably, similar to (1+1) dimensions, we also find that $\alpha < \alpha_s$ still holds when long-range temporal correlation is introduced in the most physically relevant dimension. As mentioned above, more new theoretical and numerical methods are needed for make further investigations and comparisons.

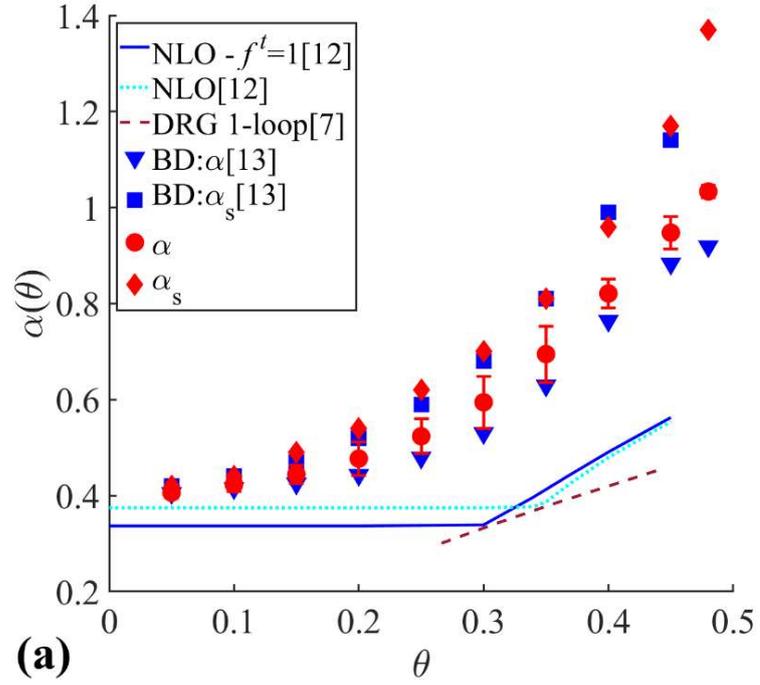

**(a)**

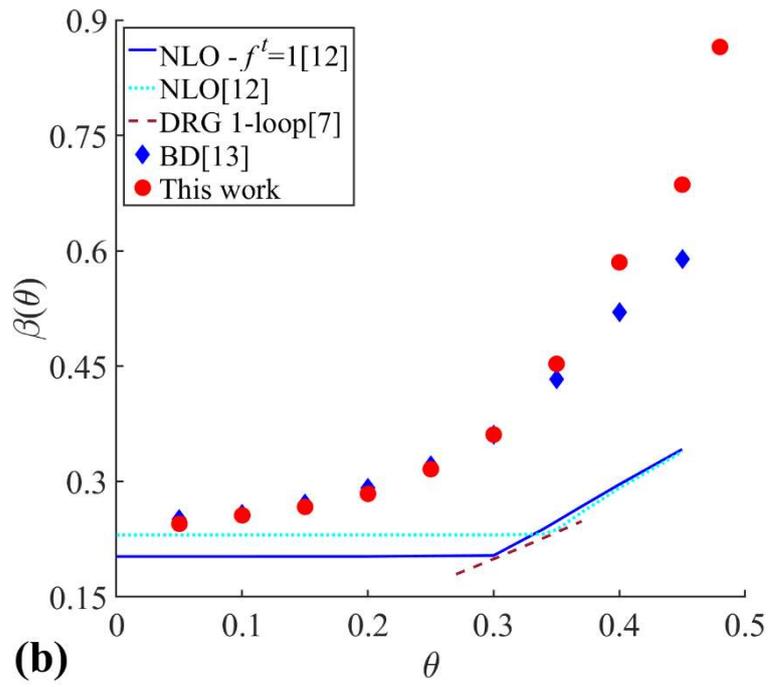

**(b)**

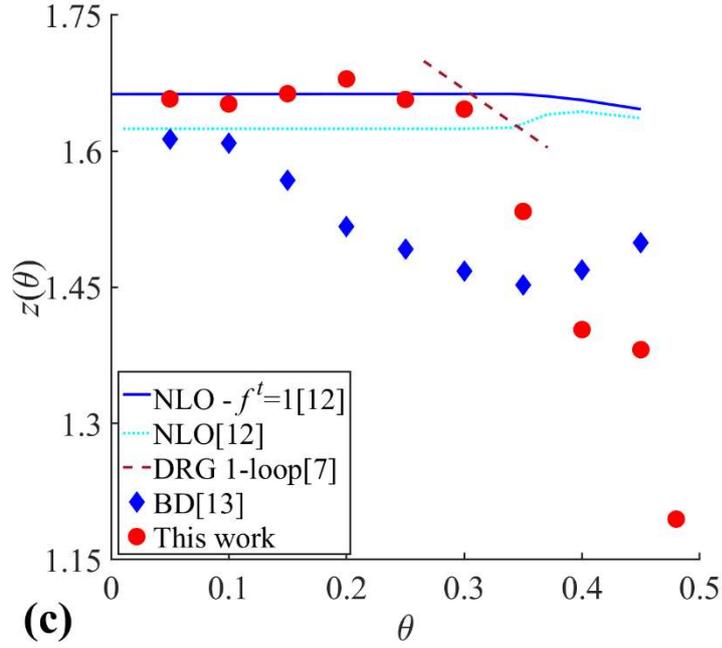

Figure 12. The values of the critical exponents in (2+1) dimensions: (a) $\alpha$ and $\alpha_s$, (b) $\beta$, c) $z$. The existing theoretical predictions and numerical results are also provided for a quantitative comparison.

We find that the critical exponents vary continuously, and depend evidently on the temporal correlated exponent, as shown in Figure 12. Thus, the temporal correlated KPZ system displays a family of continuously changing universality classes. Meanwhile, there also exists a crossover from self-affine to mounded pattern with increasing $\theta$, as shown in Figure 7. Interestingly, there should exist a critical threshold $\theta_\tau$ separating the small and large $\theta$ regimes, as one can observe from $\theta=0.20$ to $\theta=0.40$ in Figure 7(f)-(g). The saturated surfaces still display self-affine, and there is no obvious mounds at $\theta=0.05$ and $\theta=0.20$. However, obvious mounds appear when $\theta=0.40$, and the full faceted pattern gradually forms at $\theta=0.48$ in the long time limit [13]. Therefore, the (2+1)-dimensional temporal correlated KPZ system has similar properties to that of the (1+1)-dimensional case. Recent literatures show that there exists the critical threshold ($\theta_\tau \approx 1/4$) in the (1+1)-dimensional linear Edwards-Wilkinson and nonlinear KPZ in the presence of the temporally correlated noise [11, 33]. Our results also imply that the critical threshold in the stochastic growth model driven by the temporally correlated noise is universal, which does not rely on the substrate dimension.

## 5. Conclusions

We have investigated numerically KPZ growth with long-range temporal correlations in (1+1) and (2+1) dimensions. In the (1+1)-dimensional case, our results are in line with the previous values from simulating KPZ and BD growth system driven by the long-range temporally correlated noise, and the scaling exponents are consistent with SCE predictions for the small $\theta$ regime, and are in agreement with FRG for the large $\theta$ regime. We find that surface morphologies are obviously changed with temporal correlations, and the global and spectrum roughness exponents differ with each other, which implies that the dynamic growth in the KPZ subject to long-range temporally correlated noise display nontrivial scaling behavior. For the (2+1)-dimensional case, we obtained the scaling exponents in the whole effective $\theta$ regime, and when $\theta \to 0$, these results obtained here are recovered to the existing numerical and theoretical predictions on the local KPZ with uncorrelated noise. When $\theta > 0$, we find that long-range temporal correlations affect evidently the scaling behavior during the early and saturated growth regimes, unfortunately, our results are not close to the existing theoretical predictions, including DRG and NPRG approaches. We also notice that there exist somewhat differences between the BD and KPZ systems when the two models are in presence of the long-range temporal correlations, although the BD model with uncorrelated noise has long been regarded as the same universality class with the local KPZ equation. We conjecture about the difference from that one needs binarizing the correlated noise into integers in the modified BD model, while does not introduce this binarization in the discretized version of the temporal correlated KPZ equation.

As a typical local growth system, the KPZ with no correlated noise exhibits normal scaling behavior. However, strong temporal correlations in both (1+1) and (2+1) dimensions significantly changes surface morphology and scaling exponents. In the presence of long-range temporal correlations, the growth surface displays nontrivial dynamic scaling when $0 < \theta < 0.5$, for example, $\alpha < \alpha_s$, but $\alpha_{loc}$ exhibits non-universal scaling exponent, and thus the local scaling is not evident within small window size. It is not yet fully understood why the scaling properties of the temporal correlated KPZ system are beyond the existing theoretical predictions [22]. Thus, further exploration and research of the temporal correlated growth models is still necessary. We expect that our results can potentially be of interest for other problems involving the same universality class of the temporal correlated KPZ growth and other related non-equilibrium system.

## Acknowledgements

We would like to thank Yueheng Lan for useful discussions and critical reading of the manuscript. This work was supported by Natural Science Foundation of Jiangsu Province (No. BK20180637); National Natural Science Foundation of China (NSFC) (No. 11804383).